\documentclass[12pt,preprint]{aastex}
\usepackage{natbib,epsfig,chngcntr}
\tighten

\newcommand{\Planck}{{\it Planck}}

\begin{document}

\slugcomment{Accepted by the Astrophysical Journal; v.30 Nov 2015; arXiv:1510.00126} 
\shorttitle{Evidence for Alternate Universes?}

\title{Spectral Variations of the Sky: Constraints on Alternate Universes}
\author{R. Chary\altaffilmark{1,2}}

\altaffiltext{1}{Infrared Processing and Analysis Center, MS314-6, California Institute of Technology, Pasadena, CA 91125; rchary@caltech.edu}
\altaffiltext{2}{Visiting Associate, Smithsonian Astrophysical Observatory, 60 Garden Street, Cambridge, MA 02138}

\begin{abstract}
The fine tuning of parameters required to reproduce our present day Universe suggests that our Universe may simply be a region within an eternally inflating super-region. Many other regions beyond our observable Universe would exist with each such region governed by a different set of physical parameters. Collision between these regions, if they occur, should leave signatures of anisotropy in the cosmic microwave background (CMB) but have not been seen. We analyze the spectral properties of masked, foreground-cleaned maps between 100 and 545 GHz constructed from the Planck dataset. Four distinct $\sim2-4\arcdeg$ regions associated with CMB cold spots show anomalously strong 143 GHz emission but no correspondingly strong emission at either 100 or 217 GHz. The signal to noise of this 143 GHz residual emission is at the $\gtrsim$6$\sigma$ level which reduces to $3.2-5.4\sigma$ after subtraction of remaining synchrotron/free-free foregrounds. We assess different mechanisms for this residual emission and conclude that although there is a 30\% probability that noise fluctuations may cause foregrounds to fall within 3$\sigma$ of the excess,  there is less than a 0.5\% probability that foregrounds can explain all the excess. A plausible explanation is that the collision of our Universe with an alternate Universe whose baryon to photon ratio is a factor of $\sim$4500 larger than ours, could produce enhanced Hydrogen Paschen-series emission at the epoch of recombination.  Future spectral mapping and deeper observations at 100 and 217 GHz are needed to mitigate systematics arising from unknown Galactic foregrounds and to confirm this unusual hypothesis.
\end{abstract}


\section{Introduction}

Following a hot Big Bang, as the Universe cooled, the electrons and protons recombined at redshifts between $z\sim1500$ and 1000. The photons last decoupled from the particles at $z\sim1050$
at which point the Universe became transparent. This process of recombination, causes bound-bound transitions of the electrons within the energy shells
of Hydrogen and Helium as well as free-bound transitions, both of which produce line and continuum emission.
Detailed modeling of the recombination line emission has been presented and their contribution as distortions to the cosmic microwave background spectrum
at GHz frequencies previously calculated \citep{Haimoud2013, Haimoud2011, JARM2008, Chluba2006} . From these redshifts, the recombination line emission of the Balmer and Paschen series 
is redshifted to the frequency range $100-1000$ GHz, conveniently observed by the \Planck\ satellite.

The \Planck\ satellite
\citep{Overview2014} was launched on 14 May 2009 and observed the sky stably and continuously from 12~August 2009 to 23~October 2013.  \Planck's scientific payload 
included the High Frequency Instrument \citep[HFI;][]{Lamarre2010} whose bolometers cover bands centered at 100, 143, 217, 353, 545, and $857\,$GHz.  \Planck\ imaged the whole sky twice per year, with a combination of sensitivity, angular resolution, and frequency coverage never before achieved.  The HFI instrument ran out of liquid Helium cryogen in Jan 2012 which allowed for almost five full sky surveys to be
performed. The multi-wavelength coverage has allowed separation of the individual components of emission
with spectacular precision \citep{HFIBandpasses, HFICO}. 
%These include the temperature anisotropies of the cosmic microwave background, foreground discrete sources such as infrared bright star-forming galaxies, radio galaxies and interstellar medium
%emission. The \Planck\ bandpasses also straddle energetically important CO lines which has allowed measurement of the Galactic CO emission, even at high Galactic latitudes \citep{HFIBandpasses, HFICO}. 
However, \Planck\ (and {\it WMAP}) are not absolutely calibrated missions, and as such are insensitive to spectral distortions of the CMB blackbody, i.e. the monopole, which was exquisitely measured by the FIRAS
instrument on COBE \citep{Mather1994}. 

The cold spots in the CMB correspond to the deeper potential wells produced by quantum fluctuations.
Since the recombination rates in over dense regions of the Universe could be higher than those in under dense regions by the clumping factor of the plasma,
one can search for the recombination line signature in the CMB cold spots, despite the lack of spectral constraints on the monopole by \Planck.
By convolving the simulated recombination line spectrum \citep{JARM2008, Chluba2007, Chluba2006} through the \Planck\ bandpasses, it is easy to estimate that the spectral distortion due to line emission is several orders of magnitude
below the noise threshold of the \Planck\ all sky maps (Table \ref{tbl1}). Based on
the relative signal-to-noise ratios, it can be seen that the recombination line signature would be the strongest in the \Planck\ 143 and 353 GHz bands.

The fine structure constant and baryon density of our Universe makes it challenging to detect such recombination line emission with the existing data (Table \ref{tbl1}). However,
if bubbles of alternate Universes with physical properties radically different from our own \citep{Aguirre2005}, intersected in space-time with our Universe, it could potentially result in 
anisotropies in the CMB temperature fluctuations \citep{Dai2013} and an enhancement of the recombination line signature making it more observable in existing data. Although temperature anisotropies in the CMB fluctuations have been found to be consistent with an isotropic Universe \citep{Isotropy2015}, these analysis have relied on the spatial statistics of the resultant CMB and
have not utilized the spectral variation of the CMB.  For example, using WMAP data, it was argued that the number of colliding bubbles is $<1.6$ at 68\% confidence \citep[e.g.][]{Feeney2011}.
Here, we attempt to constrain the recombination line signature through spectral analysis of the CMB distortions in the \Planck\ bands and through this
analysis, assess the presence of Universes with properties that are different from our own. We note that the regions that result from the interaction of our Universe with other Universes need to be
larger than the Hubble horizon at the surface of last scattering; this implies than any such regions would be $\gg$1$\arcdeg$.

As has been pointed out by \citet{Desjacques2015}, contamination from foreground sources, particularly cosmic infrared background fluctuations arising from clumped populations of high-z infrared luminous galaxies could mimic the spectral variations from recombination line emission unless spectral resolutions $\gtrsim$5 are available. In addition, temperature
and emissivity variations of Galactic cirrus and CO line emission could potentially introduce excess emission in individual bandpasses that would need to be measured and removed.

The recombination line signature may be strongest in the CMB cold spots; we therefore undertake a differential analysis where we measure the strength of 
foreground emission in the direction of CMB hot spots ($\Delta$T$_{\rm CMB}>0$\,K) to subtract it
in the direction of the CMB cold spots ($\Delta$T$_{\rm CMB}<0$\,K). Since the CMB temperature anisotropies are to first order, uncorrelated with the presence of foreground emission such as galaxy clusters (responsible for the
Sunyaev-Zeldovich effect), cirrus emission and CIB fluctuations, the CMB hot spots should provide a robust template for eliminating all foregrounds. Although second-order effects like the late-time
Sachs-Wolfe effect can result in a correlation between CMB cold spots and foreground overdensities, these should not produce a spectral variation across the Planck bandpasses. We describe the
process of foreground cleaning in section 2.

After cleaning the individual frequency maps of CMB and foregrounds, we should have noise dominated maps with the only residual emission arising from recombination line emission from $z\sim1000$.
We study the properties of the smoothed residual maps in section 3 paying particular attention to the 143 and 353 GHz bands where recombination line emission is expected to be the strongest. 
In section 4, we assess the significance of excess seen in the residual emission and provide various scenarios to explain this 
excess. We summarize our conclusions in section 5.

\section{Removing the CMB and Interstellar Medium Foregrounds from Frequency Maps}

We utilize the publicly released (PR2) \Planck\ full-mission, full-sky, HEALPix-format, frequency maps \citep{Healpix}
between 100 and 857 GHz. The 857 GHz frequency map serves as a template for Galactic interstellar medium emission which
is a significant contributor to the emission at high latitudes at the other frequencies 
\citep{CompSep2015}.
We also utilize the Commander and SMICA component-separated maps of the cosmic microwave background (CMB) emission \citep{CMBmaps2015}. 

As a first step, we mask out sources and Galactic plane emission.
For this, we use as a starting point the conservative
confidence mask generated for CMB component separation to mask the Galactic Plane; this is called the ``UT78 mask". The mask is extended to 
exclude regions whose Galactic latitude is less than 20$\arcdeg$.
As a further extension, galaxy clusters detected through the Sunya'ev-Zeldovich from the ``SZ-Union'' catalog are masked out to a radius of 10$\arcmin$.
Finally, for each frequency, we also mask sources from the corresponding, high-confidence, \Planck\ catalog of compact sources \citep[PCCS2;][]{PCCS2} out 
to a radius of the full-width half-maximum (FWHM) at that frequency.  The source mask used for the CMB map is the product of the masks for the 100 to 353 GHz channels since those
are the frequencies where the CMB is significant.

We convert the frequency maps at 545 and 857 GHz to thermodynamic temperature Kelvin$_{\rm CMB}$ units using the calibration factors provided in \citet{Calib2015}; the lower frequency maps
are already in these units. We then subtract the CMB map resulting from the SMICA component separation technique,
from the individual frequency maps; this
yields a CMB-free map at all frequencies, although the contribution from the CMB at 545 and 857 GHz is negligibly small. We note that since the CMB is constant across all frequencies
in Kelvin$_{\rm CMB}$ units, no frequency dependent scaling factor is required. We then  degrade the spatial resolution of the masked, CMB-free maps to 13.7$\arcmin$ (NSIDE=256).

We next subtract out foreground sources of emission, the dominant among which are the Galactic interstellar medium and the fluctuations due to the integrated emission from unresolved sources
that contribute to the cosmic infrared background. For this purpose, we convert the CMB-free maps from the earlier step, back into MJy/sr units, using the calibration factors provided in \citet{Calib2015}.
Since the properties of the interstellar medium vary on relatively small angular scales, we use a differential technique whereby the properties of
the Galactic foregrounds at the location of the CMB hot spots within 1.5$\arcdeg$ of a CMB cold spot in the CMB map 
are used to estimate the level of foreground emission at the location of the cold spot. As a consistency check, we repeat the analysis using a 0.5$\arcdeg$ search radius as well.
For each pixel corresponding to a positive CMB temperature fluctuation,
we measure the ratio of intensities at that pixel, between each frequency and the 857 GHz CMB-free map. This is effectively the ratio of Galactic foreground emission
at a particular frequency to the ISM emission at 857 GHz. If the ratio is negative, due to noise, the corresponding pixel is masked out and neglected from the analysis. 
This yields ratio maps between each frequency and
857 GHz, with the ratio spatially varying across the sky due to variation in the color temperature of the foregrounds
(See Appendix A). The median intensity ratios, relative to 857 GHz in MJy/sr units, are 0.0024, 0.0078, 0.033, 0.14, 0.41 at 100, 143, 217, 353 and 545
GHz respectively. The standard deviations of the ratios across the sky are 0.0012, 0.0020, 0.0067, 0.020, 0.043 at these frequencies respectively. 
The standard deviations are not the noise in the
measurements but illustrate the range in colors of the ISM as illustrated in the all sky maps in Figure \ref{fig:ratiomaps}. For comparison, a 17.5K far-infrared color temperature blackbody 
with $\nu^{1.4}$ dust emissivity would show ratios of
0.0025, 0.0076, 0.03, 0.13 and 0.40 illustrating that the median ratios are consistent with ISM dust although with an emissivity that is somewhat lower
than the canonical value of 1.6. All numbers in this paper, are derived for analysis with the SMICA CMB map subtracted, unless explicitly stated.

For each negative CMB temperature fluctuation i.e a CMB cold spot, we then identify all pixels within 1.5$\arcdeg$ radius which correspond to a CMB hot spot and have a valid color ratio measured.
We take the median of these pixels and use that value as the colors of the foreground emission for the CMB cold spot. We scale the 857 GHz emission at that CMB cold spot pixel by this ratio
and subtract it from the corresponding pixel at each of the frequencies. This yields a map of the sky, free of CMB fluctuations and all foregrounds that were detectable at 857 GHz and provides the basis for further analysis.
%These maps have  standard deviations of 0.0092, 0.0117, 0.00857, 0.0119, 0.0267 and 0.038 MJy/sr at 100 through 545 GHz respectively. 
These maps have  standard deviations of 0.0023, 0.0020, 0.0043, 0.012 and 0.025 MJy/sr at 100 through 545 GHz respectively. We note that using a smaller radius of 0.5$\arcdeg$ for
foreground subtraction results in noise values which are 1.06, 0.98, 0.92, 0.9, 0.86 times these values. This is not surprising - at frequencies where the ISM dominates, using a small radius accounts
for structure in the ISM and results in cleaner foreground subtraction while having $\sim$10\% fewer pixels left for analysis.
These map products are then rebinned to lower spatial resolution corresponding to pixels of 1.83 and 3.67$\arcdeg$, while ignoring the pixels that are masked. It is worth noting that the footprint of the
low spatial resolution pixels overlaps both CMB hot and cold spots but since the residual at the location of the hotspots at all frequencies is zero, the low resolution maps are a measure of the residual 
emission at only the location of the CMB cold spots.

Upon publication, all map products as a result of this analysis will be publicly available at this website for community analysis\footnote{\url{http://www.its.caltech.edu/$\sim$rchary/multiverse/}}.

\section{Spectral Variations of the Sky}

After subtracting out foregrounds and masking pixels which have detectable sources, the rebinned, source-masked, residual maps are free of spatial structures and should be noise dominated.
Noise maps are generated by propagating the measurement uncertainties through the entire analysis procedure. For the CMB subtraction, we simply use the difference in the CMB maps
generated from the \Planck\ half-ring maps as a tracer of the uncertainty in fitting for the CMB. 
We note that due to the \Planck\ scan strategy, the range of measurement
noise values in the maps spans a wide range, $>$40, and depends on the frequency\footnote{An earlier version of this analysis neglected the spatial variation of the noise and obtained incorrect results}. The residual and noise maps are shown in Figure \ref{fig:snrmaps}.
Dividing the residual map with a noise map yields a low resolution signal to noise map. 
The properties of the signal to noise maps and residual signal are discussed in the Appendix.

Histograms of the signal to noise clearly show that although the median of the maps are
zero at 217, 353 and 545 GHz, at 100 and 143 GHz, they are $\sim$2$\sigma$ offset from zero. The median intensity has a value of 1.43$\pm$0.78 and 1.04$\pm$0.9 kJy\,sr$^{-1}$ at these two frequencies respectively
with the standard deviation measured over the unmasked sky. It
can be fit with a I$_{\nu}\propto\nu^{-0.7\pm0.3}$ spectrum indicating that it is residual high latitude synchrotron emission which is likely uncorrelated with the diffuse ISM that was detected at 857
GHz and which was used to subtract the foreground ISM. 

Both synchrotron and free-free emission have well-defined spectral indices unlike thermal dust emission which can span a wide range of color temperatures. However,
the noise in the \Planck\ 100 GHz maps is substantially higher than that at 143 GHz (Table 1) which would result in a noisy residual map if we were to 
subtract this synchrotron emission using the 100 GHz residual map as a template, without knowing the spatial variation in the spectral index of emission. As a result, we first 
identify pixels which have signal to noise ratios (SNR) at 143 and/or 353 GHz greater than 5 since these are the band passes where the intensity of the recombination lines from $z\sim1000$
is thought to be the strongest.  We then sub-select among these, the pixels whose residual emission at 143 GHz is greater than 2$\sigma$ above the residual flux density at 100
and 217 GHz. That is, since we are searching for line emission in the \Planck\ bandpasses, we require $(I_{143}-\sigma_{143})>(I_{100}+\sigma_{100})$ and $(I_{143}-\sigma_{143})>(I_{217}+\sigma_{217})$. Furthermore, due to the possibility of residual ISM contamination, we require that the ISM values lie in the lowest third of the observed range i.e.
pixels with 143 GHz SNR$>$5 fall in the range 0.8 to 12.5 MJy\,sr$^{-1}$ in the 857 GHz frequency map and we place an upper limit of 4 MJy\,sr$^{-1}$ for the allowable range of intensity.

\begin{figure}
\epsscale{0.5}
\plotone{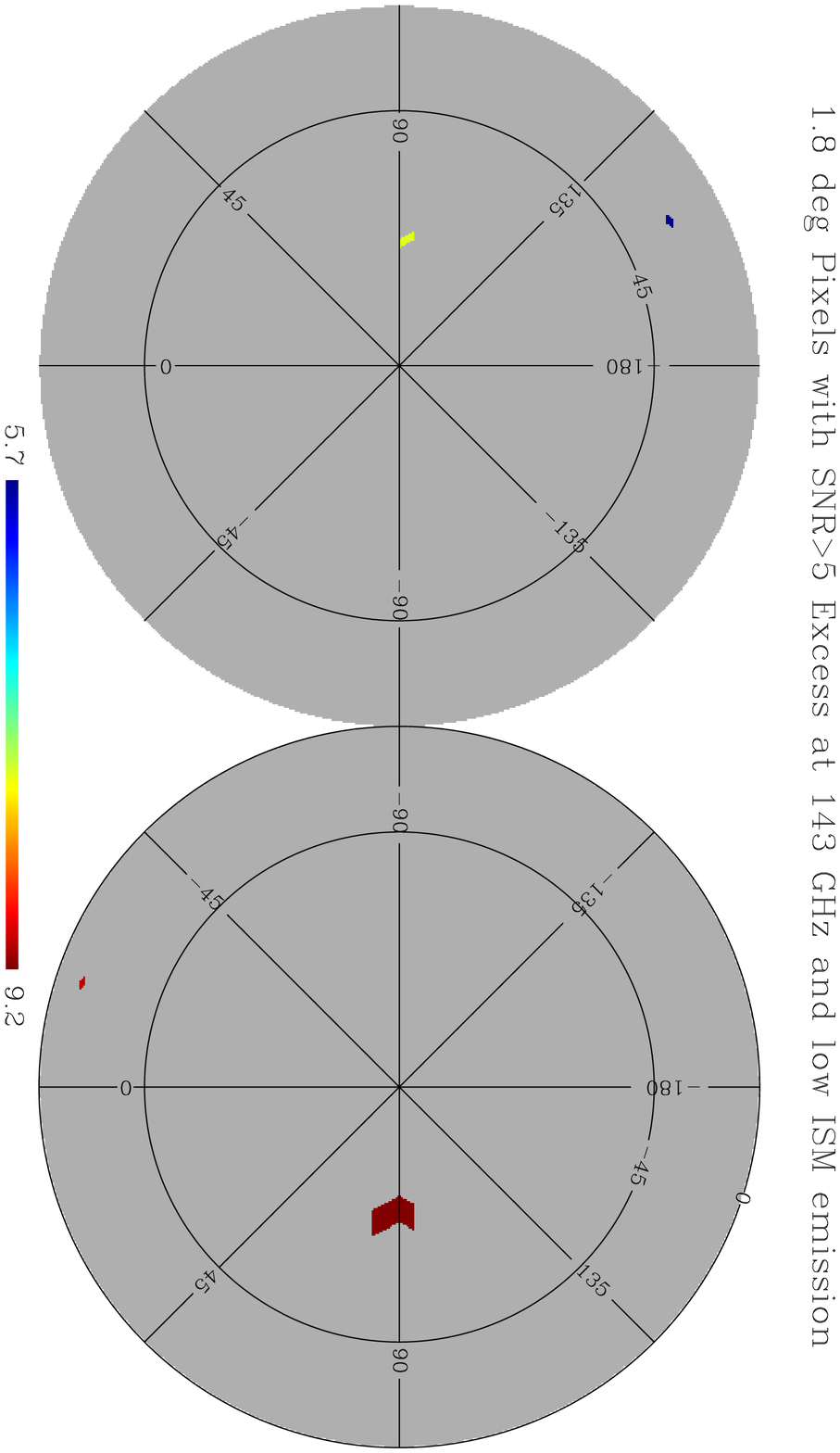}
\caption{
Galactic orthographic projections of the entire sky showing the 5 positions on the sky which have SNR$>5$ excess at 143 GHz after foreground cleaning,
and whose residual 143 GHz intensity is more than
2$\sigma$ above the 100 and 217 GHz intensities. All the points have 857 GHz ISM emission less than 4 MJy\,sr$^{-1}$. The points are color coded by 143 GHz signal to noise ratio.
The left circle shows the Northern hemisphere while the right circle shows the Southern hemisphere with the Galactic equator around the circumference of the circles.
The pixel scale is 1.8$\arcdeg$. Exact coordinates are given in Table 2.
If this excess arises from enhanced recombination
line emission at $z\sim1000$, it would argue in favor of a collision with an alternate Universe with a 
higher baryon to photon ratio than our own. A Monte-Carlo
analysis indicates that this excess is significant at the $>$5$\sigma$ level 44\% of the time.
A simpler explanation would favor noise spikes on top of a combination of Galactic synchrotron and free-free emission.
}
\label{fig1}
\end{figure}

\begin{figure}
\epsscale{0.6}
\plotone{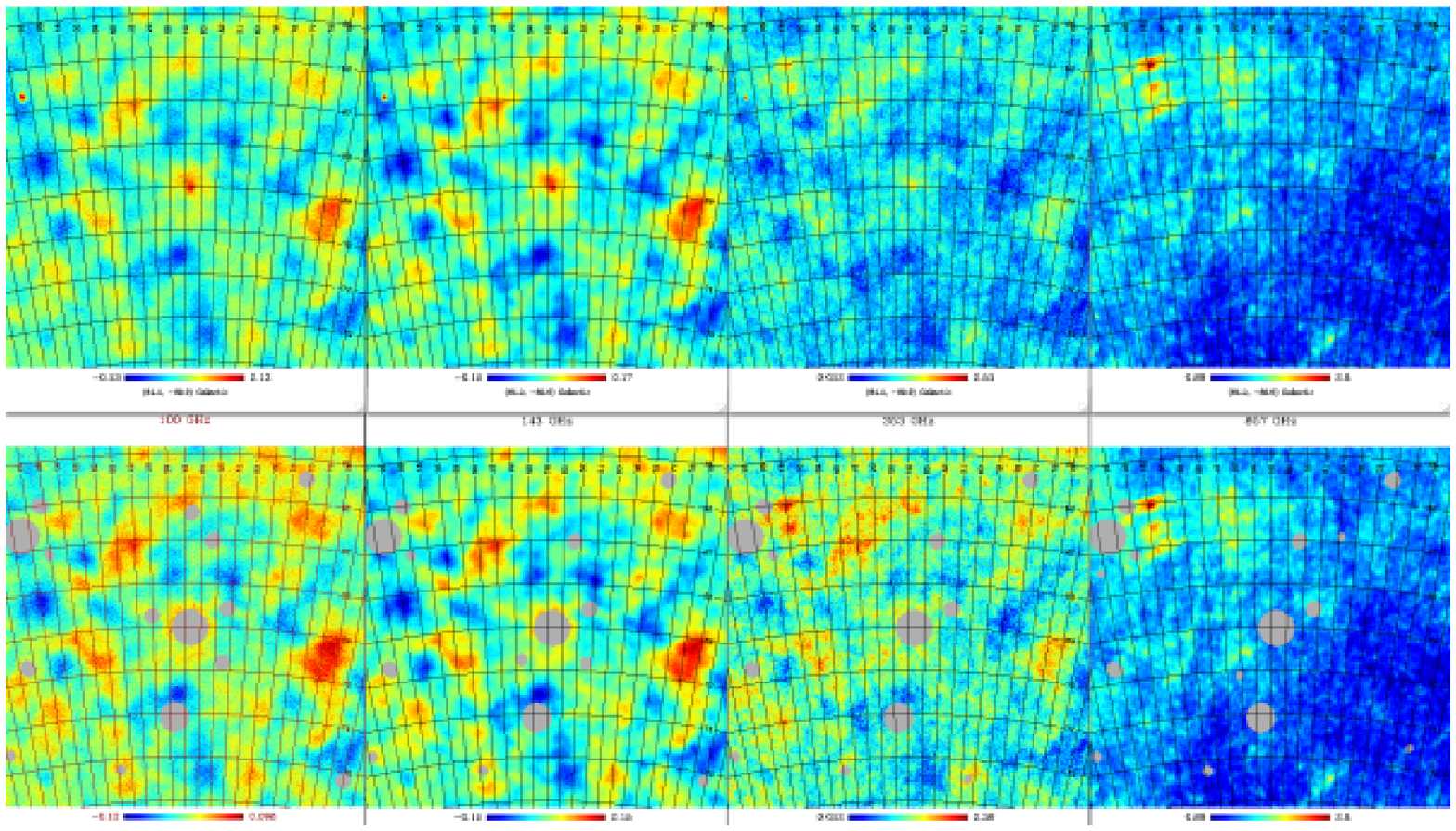}
\plotone{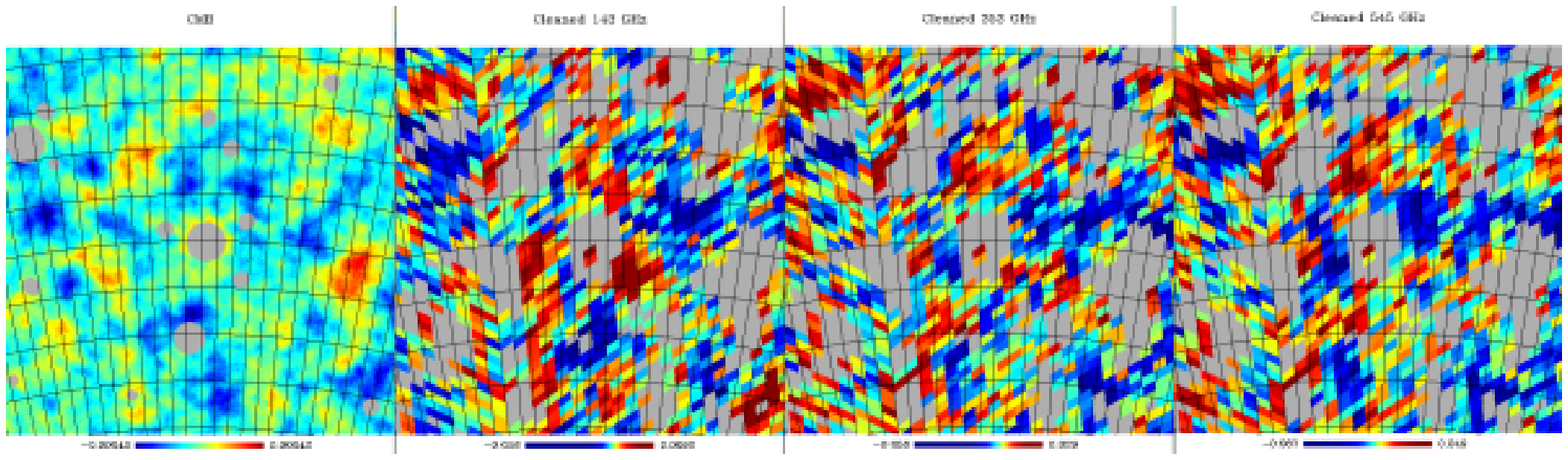}
\plotone{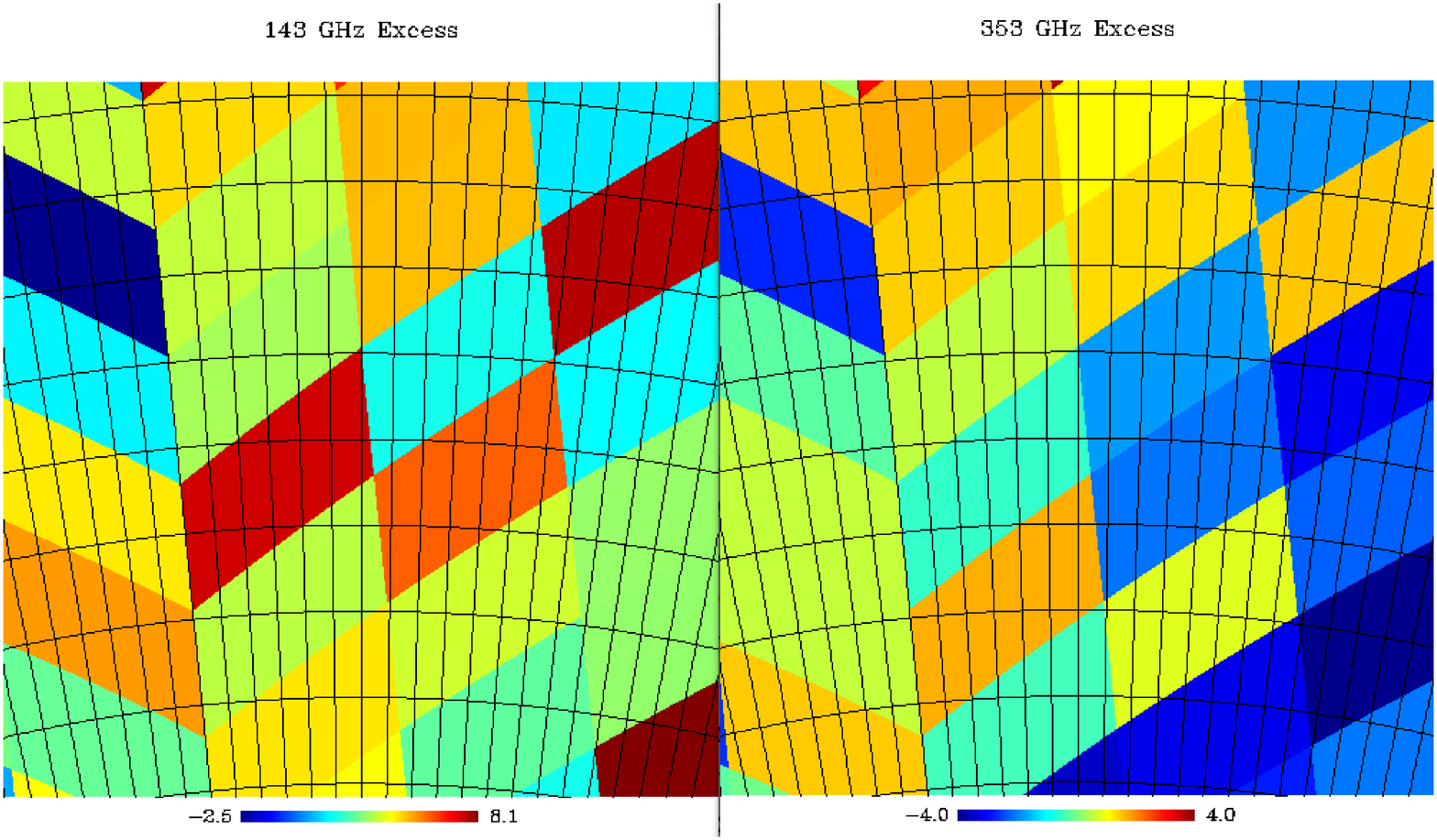}
\caption{
Multi-panel image showing the evolution of the data processing from the frequency maps to the final result at the location ($l,b=84\arcdeg, -69\arcdeg$).
The top row shows the 100, 143, 353 and 857 GHz \Planck\ frequency maps at a scale of 1$\arcmin$ per pixel, subtending an angle of 8.33$\arcdeg$. The second row shows the same data folded through the foreground source mask.
The intensity scale is chosen such that blue is low intensity and red is high intensity. The third row shows the hot and cold spots corresponding to the component separated CMB map, along with the 143 GHz,
353 GHz and 545 GHz residual maps after CMB and foreground subtraction. Note that the residuals are zero at the location of the CMB hot spots by design.
The final row shows the smoothed and averaged, excess SNR maps at 143 and 353 GHz with each pixel
subtending 1.8$\arcdeg$. The frequency maps have an intensity scale shown in MJy\,sr$^{-1}$. The central regions of the maps appear to show excess 143 GHz
emission at the location of CMB cold spots that do not correspond to regions of strong ISM emission i.e. $\ll4$ MJy/sr at 857 GHz.
}
\label{fig2}
\end{figure}

After eliminating edge effects, there are a total of 5 regions we find which meet these criteria; three are based on the analysis using the SMICA map while two are based on the analysis using the Commander
CMB map (Figure \ref{fig1}). They are listed in Table \ref{tbl2}. 
The 143 GHz intensity in these regions is $\sim1.3-3.2\times10^{-3}$\,MJy/sr in these regions with an uncertainty of $0.14-0.4 \times10^{-3}$\,MJy/sr
depending on the area over which the signal is averaged (Table 2). 
Of these, the only region which has structure on both 1.8$\arcdeg$ and 3.6$\arcdeg$ spatial scales is the one at $(l,b)=(84\arcdeg,-69\arcdeg)$ (Figure \ref{fig2}).
The 143 GHz residual intensity in this region is $1.3\pm$0.14 kJy\,sr$^{-1}$ at the 3.7$\arcdeg$ scale and $2.5\pm0.34$\, kJy\,sr$^{-1}$ at the 1.8$\arcdeg$ scale.

Although the statistical significance of this excess is high, the noise in the adjacent bands, 100 and 217 GHz is higher than at 143 GHz. Therefore, it is prudent to Monte-Carlo the 
foreground emission to assess the true statistical significance of this excess. We use a random number generator to scatter the measured residual intensities in the three bands
100, 143 and 217 GHz. The random numbers have a standard deviation which is the same as the measurement uncertainty at each frequency (Table 2). 
For each of the 1000 iterations, we fit the randomized 100 and 217 GHz values
with a power-law to predict the 143 GHz intensity. We compare this interpolated intensity with the correspondingly randomized 143 GHz intensity from the Monte-Carlo. We then find the percentage of
trials where the scattered intensity is $>$5$\sigma$ or $>$3$\sigma$ of the interpolated 143 GHz intensity. The percentage of such trials is listed in Table 3. 

At $(l,b)=(84\arcdeg, -69\arcdeg)$, the most promising region of the 5 selected regions where the 143 GHz excess extends on scales of $\sim$2-4$\arcdeg$, 
we find that the probability that the excess is significant at the $>3\sigma$ level is 76\% while the probability that the excess is significant at the $>5\sigma$ level is 
44\%. Thus, although the excess emission at 143 GHz in those regions is $\gg$5$\sigma$ in terms of measurement uncertainty, due to contamination from synchrotron emission
and the large uncertainty in the adjacent bands at 100 and 217 GHz, which makes for uncertainty in fitting the residual foreground emission, the statistical significance of the excess is
substantially reduced.

Also shown in Table \ref{tbl3}, is the signal to noise of the 143 GHz excess (SNR$_{excess,143}$) after a power-law (I$_{\nu}\propto\nu^{\alpha}$)
to the synchrotron or free-free foreground is fit to the measured 
residuals at 100 and 217 GHz and interpolated to 143 GHz. This simply shows how far above the residual foregrounds the 143 GHz excess is and does not take into account the
measurement uncertainty at 100 and 217 GHz which we have factored in the Monte-Carlo analysis\footnote{In the Appendix, we repeated this analysis while including the \Planck\ measurement
at 30 GHz for a stronger constraint on the foregrounds. There was little impact on the conclusion.}.

Finally, in order to confirm that the appearance of excess at 143 GHz is a feature of pixels that are associated with CMB cold spots, we repeated the exercise in reverse. 
That is, for each pixel corresponding to a negative CMB temperature fluctuation,
we measure the ratio of intensities at that pixel, between each frequency and the 857 GHz CMB-free map. Then for each CMB hot spot, corresponding to an under density at the epoch of recombination,
we then identify all pixels within 1.5$\arcdeg$ (or 0.5$\arcdeg$) radius which correspond to a CMB cold spot and have a valid color ratio measured.
We take the median of these pixels and use that value as the colors of the foreground emission for the CMB hot spot to obtain a map which should contain all foregrounds removed at the location of the CMB
hot spots. These maps are similarly rebinned to lower spatial resolution while ignoring the pixels that are masked. We find that these map products do not contain any significant
evidence for excess emission at 143 GHz indicating that under densities probed by CMB fluctuations do not correspond to residual excess emission. We do find evidence
for significant excess at 353 GHz but the strength of the emission correlates well with the far-infrared color temperature of the foreground emissions in the sense that regions of cooler dust with higher 545/857 intensities
show stronger 353 GHz residual emission. This implies that at 353 GHz, the contamination from cold ISM which has substantial structure on small angular scales, is too large to make detection of epoch of recombination signals possible. At 143 GHz however, as demonstrated above, there is significant potential to look for spectral anomalies in the direction of the CMB cold spots since
the dominant foreground residual is synchrotron emission which has a well-constrained spectral shape.

%For example, all the regions with 353 GHz residual excess have 545/857 GHz intensity ratios of $0.34-0.42$ and with 857 GHz intensities ranging from 1.9 to 16.6 MJy/sr
%indicating that much of the residual excess in this analysis is because of variation in dust temperatures on small angular scales. 

\section{Interpretation of the Spectral Variations}

We have cleaned the \Planck\ frequency maps of CMB and foreground interstellar emission between 100 and 545 GHz. We expected the spectral signal at the location of the CMB cold spots
after smoothing to several degree angular scales to have a mean of zero and be noise dominated. Instead we have found the intensity of the residual signal at 100 and
143 GHz in the direction of CMB cold spots, is $\sim$2$\sigma$ offset from zero over the entire sky (Figure \ref{fig:residualhist}). Comparison with the CMB emission or strength of 
ISM emission shows no evidence of a correlation (Figures \ref{fig:cmbcorr} and \ref{fig:ismcorr}). However, the spectral index of the mean residual emission between 100 and 143 GHz can best be
explained by a I$_{\nu}\propto\nu^{-0.69}$ synchrotron spectrum suggesting that it is foreground Galactic emission.

After accounting for spatial variation of the synchrotron emission using the 100 and 217 GHz residual maps,
we find that five separate regions, of which two are adjacent, show unusually strong residual emission at 143 GHz which is in excess of emission at 100 and 217 GHz. 
We undertook a Monte-Carlo analysis whereby we accounted for residual foreground emission by fitting the 100, 143 and 217 GHz residual intensity values with a single power law.
 The $\chi^{2}$ distribution reveals that the probability that all of this emission is due to synchrotron or free-free foregrounds is small. It is less than 0.01\% for three of 
the regions and 0.3\% for the region at $(151.9\arcdeg, 31.4\arcdeg)$. However, we find a probability of 35\% for the region at $(86.8\arcdeg, -69.4\arcdeg)$
being fit by residual free-free emission.

Conversely, we use the Monte-Carlo to assess how large the 143 GHz excess is relative to the emission at 100 and 217 GHz (Table 3). We find that two
of the regions are 5$\sigma$ in excess above the interpolated 100 to 217 GHz spectrum $\sim$40\% of the time and are 3$\sigma$ in excess almost 70\% of the time.
Although this is intriguing, the uncertainties in the adjacent bands at 100 and 217 GHz are too large to provide definitive evidence for a 143 GHz excess.

It is clear from the absence of residuals at higher frequencies that the excess is not arising due to thermal dust emission from the ISM (Figure \ref{fig:ismresid}). This is also ruled out by the differential method
undertaken to remove the foregrounds, whereby we measure the colors of the foregrounds relative to 857 GHz at the location of the CMB hot spots to subtract the emission from
the CMB cold spots. Calibration errors are also unlikely since the calibration accuracy in these bands is 0.1\% \citep{Calib2015}.

A possibility based on the spectrum of residual emission is there may be a faint, Gigahertz-peaked radio source that may be unmasked or leaking flux density outside the mask.
However, comparison with sources in the \Planck\ catalog of compact sources reveals that all three radio sources in the vicinity of the region at $(83.6\arcdeg, -69.4\arcdeg)$ are
synchrotron dominated sources with a spectral index $\alpha=-1.4$ where $F_{\nu}\propto\nu^{\alpha}$. The sources in the vicinity of the region at $(93.2\arcdeg, 69.4\arcdeg$)
do show a peak in the spectrum at 143 GHz with a peak flux density of 500 mJy. However, the residual emission is 3.2 kJy\,sr$^{-1}$
resulting in a total flux of 3.2 Jy, much higher than the brightness of the source itself. We therefore do not think that emission leaking out from behind the source mask can
account for the spectrum of this region.

Thus, if these anomalies are not extreme noise fluctuations, as analyzed above, the most likely possibilities are line emission from the ISM or cosmological/extragalactic line emission. 
We assess each of these in turn.

\subsection{Alternate Universes?}

We have found evidence that some of the CMB cold spots appear to have temperature decrements in the \Planck\ 143 GHz bandpass that are not as large as at other frequencies. These decrements, which can be
detected after removing the foregrounds from the frequency maps, seem to occur primarily
on angular scales of $1.8-4\arcdeg$ which is larger than the Hubble horizon at the epoch of recombination. Although we cannot conclusively rule out continuum emission of an unknown form as responsible
for this excess, the absence of significantly strong excess at adjacent frequencies, and the differential method used to remove the foregrounds, suggests the contribution of line emission that is preferentially
associated with the CMB cold spot. 

If the lines indeed arise from the epoch of recombination, it is a factor of $\sim$ 4500 enhancement of the recombination line emission compared to that expected from the models. 
Specifically, it would be the Paschen series Hydrogen recombination lines from $z\gtrsim$1000 that is dominating the 143 GHz excess.
Although the line in the rest-frame is emitted at 1.87\,$\mu$m, due to the extended recombination history, in the observers frame, it spans $\sim$50 GHz in frequency. 
The bandwidth of the 143 GHz bandpass is 45.76 GHz and therefore 2.5\,kJy\,sr$^{-1}$ of residual emission at 143 GHz 
corresponds to a line flux of 1.1$\times$10$^{-12}$ W\,m$^{-2}$\,sr$^{-1}$ which is about
a factor of 4500 larger than expected from 
 fiducial models for recombination line strengths from $z\sim1000$. The clumping of baryons within the potential wells corresponding to the CMB cold spots cannot be this
 significant as the baryons are still thermalized, although that would be one plausible way to enhance recombination line emission. Another possibility is that the baryon
 density itself is enhanced by a large factor.  

There is no corresponding
anisotropy signal seen in the CMB temperature data. Although hints of asymmetry in the CMB power have been found with both \Planck\ and {\it WMAP} they favor a directionality that is associated
with regions of high ISM emission - regions that are excluded in our analysis \citep{Isotropy2015, Hansen2009}. This may not be surprising since the CMB intensity arises from photons while the recombination line signature arises from baryons. Since the recombination rate is proportional to the square of the baryon density for a neutral Universe, but the number of Paschen-$\alpha$ photons 
emitted is proportional to the number of baryons, a factor of $\sim$
4500 higher baryon to photon ratio than that observed in our Universe would be required to explain the signal. Since such anisotropies do not exist in our Universe from what we know, a plausible explanation is that
collision of our Universe with an alternate Universe with such a high baryon to photon ratio may be responsible for the higher recombination rate and thereby the detected signal \citep{Aguirre2011}.

Collisions between bubbles with different physical properties are thought to be catastrophic events, as predicted by theory \citep[e.g.][]{Kleban}. They should leave an imprint on the spatial fluctuations of the CMB
and typically are though to occur prior or during inflation. Such an event would then repeat itself across a significant fraction of sky and not be localized in space as our excess regions are. Thus, it is not clear if 
this interpretation has a robust theoretical framework. For the signature that we see to be due to bubble collisions but not affecting the CMB intensity or redshift of recombination, the collisions
need to have occurred after nucleosynthesis and before matter-radiation equality when the energy density of our Universe was much higher than that of the alternate Universe. This could result
in the injection of baryons which settled in the deeper potential wells of the density fluctuations (i.e. the cold spots) and recombined at a higher rate if these baryons are not thermalized with the CMB.
Although the baryons may be recombining at a lower rate at the location of the hot spots as well, we would be insensitive to detecting this signal due to the foreground cleaning method applied.
These baryons, if they are the origin of emission, also cannot be from our Universe since they would change nucleosynthesis in certain patches of our sky and affect the CMB intensity; in particular
a higher baryon to photon ratio would result in the production of a lot more Lithium and some $^{4}$He with a decreased abundance of $^{3}$He and $^{2}$H. 

\begin{figure}
\plotone{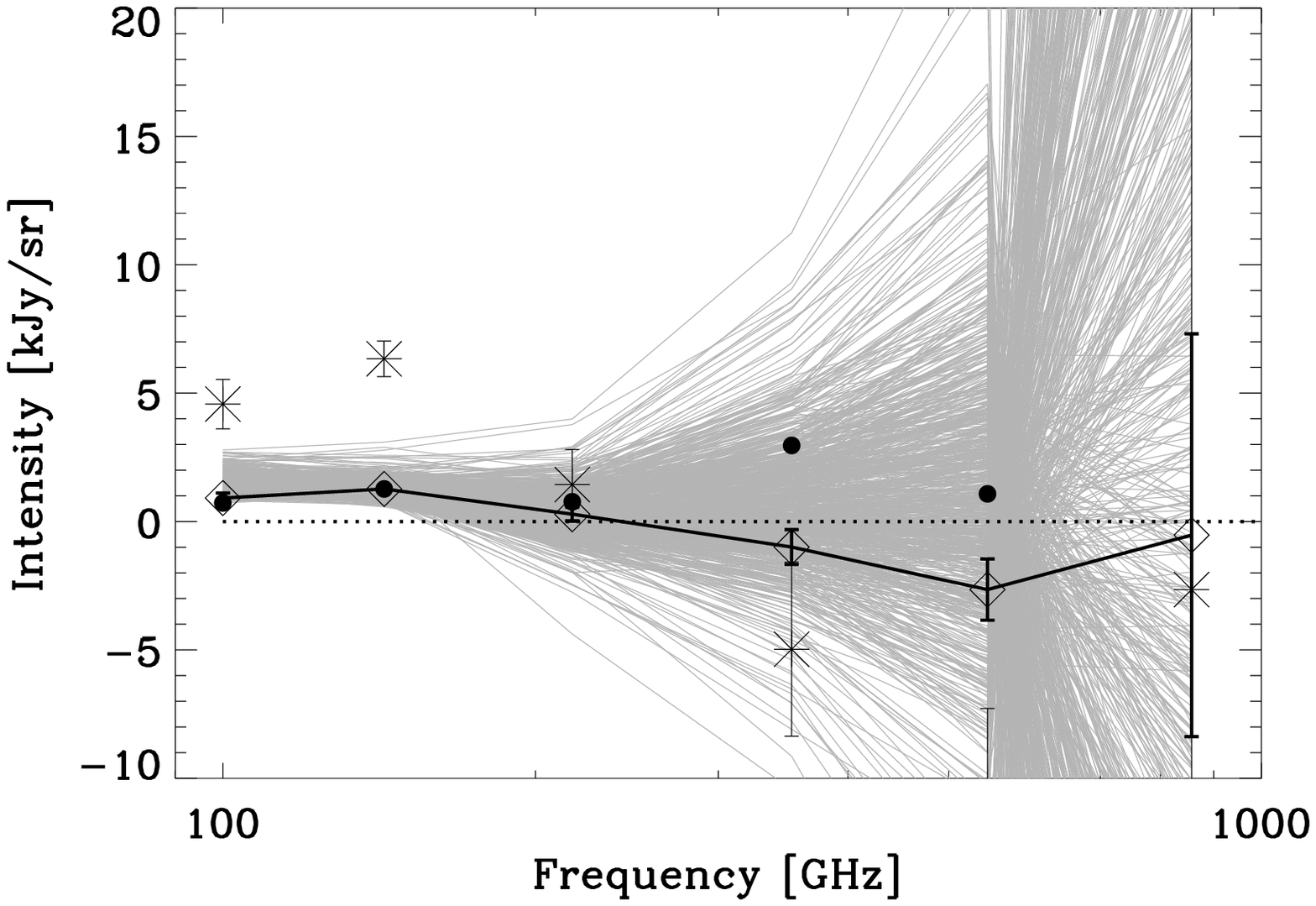}
\caption{
The thick line and diamonds indicate the spectrum of a 4$\arcdeg$ spot at ($l,b=83.6\arcdeg, -69.4\arcdeg$) where the SNR of the residual 143 GHz emission
is greater than 5 and where the 143 GHz emission is in excess of the residual at 100 and 217 GHz. The solid circles
show a simulated spectrum of the recombination line emission using the model of \citet{JARM2008, Chluba2007}, scaled by a factor of $\sim$4500.
The thin lines show the intensity spectrum of the cleaned sky between 100 and 545 GHz at pixels with 143 GHz residual SNR$>$5 and
857 GHz ISM intensity less than 4 MJy/sr which would correspond to the darkest
30\% Galactic ISM.  The CMB and ISM emission have been removed and the resultant residual
maps binned to 3.7$\arcdeg$ pixels. 
 It is clear that the shape of the spectrum at this one location is unusual, suggestive of a contribution from strongly enhanced recombination line emission while the others
 can be explained by a combination of cold thermal dust emission, synchrotron and free-free emission. The asterisks are the same residuals of the region scaled up by a factor of 5
to highlight the intensity differences between the frequencies more clearly.
}
\label{fig:spec}
\end{figure}

A convincing proof of the bubble collision hypothesis would lie in the measurement of an equivalent excess in the 353 GHz bandpass where the Hydrogen Balmer recombination lines may be present (Table 1
and Figure \ref{fig:spec}). The 353 GHz band is noisier and does show S/N$>$5 excess at many regions with low 857 GHz ISM emission. Furthermore, there are at least 2 regions
where the 143 GHz emission is stronger than the 100 GHz emission, however we find that there is evidence for significant residual at the higher frequencies, particularly 217 and 545 GHz.
In general, we find that the 353 GHz bandpass is more affected by cold residual ISM emission which is prevalent in our Galaxy and not perfectly traced by the 857 GHz emission as seen
by the range of residual intensities shown in Figure \ref{fig:spec}. Thus, we are unable to confirm the presence of excess in that band arising from the Balmer lines although we note that
the predicted excess at 353 GHz from a rescaled model, is only $\sim$2$\sigma$ offset from the measured limit. 

\citet{Chluba2009} have also demonstrated that injection of energy through dark matter annihilation
in the pre-recombination Universe may enhance the spectral distortions arising from recombination line emission. These distortions can however be broad, spanning several tens of GHz,
and depend on the redshift at which the energy is injected and the magnitude of the injected energy. They should as a result 
straddle multiple \Planck\ bandpasses which is unlike the excess that we are seeing.
Without knowing the shape of the spectral distortion across the 143 GHz bandpass, it is not possible to distinguish between these scenarios and pure recombination line emission.
However, future work will attempt to combine the low-frequency {\it Planck} channels and
{\it WMAP} data to assess the contribution of the weaker Hydrogen Brackett and Pfund-series recombination lines to the measured intensity although a first estimate suggests the low-frequency
data are too noisy
to provide useful constraints.

\subsection{Foreground Interstellar Medium?}
An alternate, less exotic possibility is that the excess emission we are seeing is due to anomalous microwave emission or line contamination from the ISM. In particular, there appears to be a region of significant ISM several degrees away from the excess (Figure 2). Despite
the extensive masking done to the foreground sources, diffuse line emission in the outskirts could contribute to the excess emission seen at 143 GHz. However, the 143 GHz does not straddle energetically
significant CO lines and it is unclear if some unknown sub millimeter lines associated with the star-forming regions may be contributing to the excess. Examples of such lines (CS3$\rightarrow$2, CH$_{3}$OCH$_{3}$) have indeed been observed previously in star-forming regions \citep{Oberg2010} and since they arise in the lukewarm envelope surrounding star-forming regions, they could be more extended that the thermal dust emission which has been masked. Since the thermal dust emission has
an intensity of $<$4 MJy/sr at this location, the ISM line emission must be over represented with respect to the thermal dust emission which is surprising since these lines are weak.

Continuum contamination from the ISM is also a possibility. The far-infrared color temperature of the emission must be as cold as $\sim$3K to fit the 143 GHz excess (Figure \ref{fig:ismresid}). This would be the coldest
dust yet detected and it would presumably be in thermal equilibrium with the CMB which is a challenge in the radiation field of our Galaxy. Such cold continuum
emission, if it arises from either residual CMB or cold ISM, would violate the limits we have at 217 GHz, be uncorrelated with the warm ISM and be
clumped on small scales preferentially in the direction of CMB cold spots which seems
like a contrived explanation.

Finally, an analysis of star-forming regions has shown that anomalous microwave emission, which arguably arises from spinning dust grains,
does not extend to frequencies as high as 143 GHz \citep{PIP15}.

Although we cannot definitively disprove the ISM contamination hypothesis, the combination of low ISM foregrounds in our regions, combined with a lack of excess at other frequencies (e.g. 100, 217 and 353 GHz) where the ISM is known to display strong CO lines, argues against this possibility.

\subsection{[CII] Emission from Galaxies?}
A third possibility for the 143 GHz excess
is line emission from clumped extragalactic sources. The strongest line arising in galaxies that could fall in this regime is the [CII] line at 158\,$\mu$m, redshifted from $10<z<15$. 
Although the strength of [CII] in reionization epoch Lyman-break galaxies is unclear,  one can make optimistic assumptions about its strength and provide an order of magnitude estimate to its contribution.
The ultraviolet measured star-formation rate density at $z\sim10$ is generally thought to be around $10^{-2.5}$\,M$_{\sun}$\,yr$^{-1}$\,Mpc$^{-3}$ \citep{Oesch2015}. Assuming that this measurement accurately
reflects the bolometric luminosity density from galaxies at these redshifts, and adopting a [CII]/L$_{\rm Bol}$ ratio of $\sim$10$^{-2}$ which is at the high end of what has been measured for low-metallicity
dwarf galaxies in the local Universe \citep{Cormier}, we find that the total contribution from [CII] line emission in the 143 GHz band is at most 1.7$\times$10$^{-14}$\,W\,m$^{-2}$\,sr$^{-1}$. This is almost
two orders of magnitude lower than the excess we measure. Thus, it is extremely unlikely that star-forming galaxies can account for the 143 GHz residual excess.

The broad bandpass of the \Planck\ HFI instrument does not allow us to use the spectral profile of the lines to definitively distinguish between these scenarios and unusual claims like evidence
for alternate Universes require a very high burden of proof. We outline pathways for the future in the next section which could help confirm the alternate Universe hypothesis.

\subsection{Future Directions}
The detection of spectral anomalies on the sky at CMB cold spots is a highly surprising result and requires further data for confirmation. It would be prudent to generate multiple realizations of the \Planck\ frequency maps with artificial spectral distortion signals injected in the sky models. Analyzing those simulated maps using the methodology adopted here will be illustrative but is beyond the supercomputing resources that currently are available within the collaboration. The only simulations that exist are multiple realizations of the CMB sky but these are less useful since we subtract the same CMB map
from each of the frequencies in our analysis. Such a simulation will help define the reliability with which \Planck\ can detect such spectral anomalies. 

LMT, GBT and ALMA spectral observations in the regions with anomalous excess can help place constraints on the contribution of weak ISM lines to the 143 GHz band intensity. 
Deep continuum observations at 100 and 220 GHz of these regions with ground-based CMB experiments
such as SPT and BICEP/Keck can place tighter constraints on the synchrotron and free-free foregrounds. However, since the redshifted Hydrogen Paschen-$\alpha$ recombination line falls in the
middle of the atmospheric oxygen absorption line at $\sim$116-120 GHz, spectral confirmation of this signal from the ground is likely to remain challenging. Space-based CMB spectral distortion observations such as with the Primordial Inflation Explorer (PIXIE) may provide the only way to conclusively test this hypothesis with special care given to sensitivity requirements such that key foregrounds are detected in a  majority of the narrow bands \citep{Kogut}. Specifically, we suggest that future CMB missions should focus on  moderate spectral resolution ($R\sim10$), multi-band spectroscopy achieving $<$0.1 kJy\,sr$^{-1}$ sensitivity,
over degree scales of sky to probe for differences in fine structure constants and recombination line strengths in the early Universe.

\section{Conclusions}

The CMB power spectrum extracted from the \Planck\ frequency maps has been shown to be consistent with a $\Lambda$CDM cosmology with a specific set of cosmological parameters measured
with unprecedented sensitivity \citep{Parameters2015}. Much of this consistency arises from precise measurements of the power spectrum on angular scales
smaller than 1$\arcdeg$.  Why these parameters are the values they are is a question that doesn't have a clear answer. One possibility is there are an infinite
set of Universes with different parameters and our Universe just happens to have the values that we measure. Searching for these alternate Universes is a challenge. One hypothesis suggests that
as each Universe evolves independently, it may collide with our observable Universe, leaving a 
signature on the signal we see. Since the CMB intensity has been shown to be isotropic, it is clear that such a collision is not seen
in the intensity of the photons. Probing the density of baryons during the epoch of recombination however provides an alternate approach. If the baryon density is higher in these alternate Universes,
recombination between baryons at redshift of $>1000$ can leave signatures of
Hydrogen recombination line emission which are redshifted into the \Planck\ bandpasses at 143 and 353 GHz. We have cleaned the \Planck\ maps at all frequencies between 100 and 545 GHz of CMB and foreground
emission to search for this recombination line signature. We find evidence for a 2$\sigma$ offset from zero in the mean residual intensity at high latitudes at 100 and 
143 GHz which is due to residual synchrotron emission in the direction of the CMB cold spots that is uncorrelated with the 857 GHz ISM template used to clean foregrounds. We also find
a 3.2-5.4$\sigma$ excess at 143 GHz on 1.8$-$3.7$\arcdeg$ angular scales, after cleaning, which arises at the location of four CMB cold spots. Since the CMB cold spots are regions of over densities in the primordial density field, it appears that enhanced
recombination line emission could arise from these over densities. However, the magnitude of enhancement is 4500 times higher than what is expected for the baryon density in our Universe.
A possible implication is that collision of our Universe with an alternate Universe that has a higher baryon density is responsible for the enhanced recombination line signature with a line flux of 1.1$\times$10$^{-12}$ W\,m$^{-2}$\,sr$^{-1}$. A more prosaic explanation using a Monte-Carlo analysis suggests that Galactic synchrotron and free-free emission scattered by measurement noise
could be within 3$\sigma$ of such significant excess 30\% of the time; yet we find that it cannot explain all the excess. We assess possible alternate reasons for this highly significant
enhancement, including the contribution of CO and other lines in the bandpass, and find that they are somewhat contrived but which can only be assessed by taking spectral data. Furthermore, if they are indeed ISM foregrounds, it does not
explain why the enhanced residual emission would preferentially occur in the direction of the CMB cold spots and not in the direction of the CMB hot spots or hide itself at frequencies
where the ISM is known to have stronger lines. 

\acknowledgements
\Planck\ (http://www.esa.int/Planck) is a project of the European Space Agency (ESA) with instruments provided by two scientific consortia funded by ESA member states and led by 
Principal Investigators from France and Italy, telescope reflectors provided through a collaboration between ESA and a scientific consortium led and funded by Denmark, and additional contributions from NASA (USA). High quality data products have been generated through extensive work by the HFI Core Team lead by J.-L. Puget and the \Planck\ Collaboration as a whole.
The author wishes to thanks A. Banday, G. Helou, M. Kleban, E. Wright, J. Chluba and J. Carpenter for feedback and discussions and G. Fazio and J. Robie for their hospitality.
This work has made use of NASA's Infrared Science Archive.

\begin{deluxetable}{ccccccc}
\tablecaption{Comparison between the expected Hydrogen and Helium recombination line signal in the \Planck\ bands and the noise properties of the frequency maps.}
\tablehead{
\colhead{Frequency (GHz)} & \colhead{100} & \colhead{143} & \colhead{217} & \colhead{353} & \colhead{545} & \colhead{857}
}
\startdata
Recombination line signal (10$^{-7}$ MJy/sr) & 1.59 &   2.79 &   1.69 &   6.52 &   2.37 &   2.15 \\
Median noise in \Planck\ maps\tablenotemark{a} (10$^{-3}$ MJy/sr) & 11  & 7.16  & 13.1  & 25.7  & 27.1  & 25.3 \\
S/N ratio (10$^{-5}$) & 1.45  & 3.89  & 1.29  & 2.53  & 0.872  & 0.851 \\
Noise in deepest part of map\tablenotemark{a} (10$^{-3}$ MJy/sr) & 2.56  & 1.74  & 3.47  & 5.24  & 5.23  & 4.97 \\
Area of deepest part of map (deg$^{2}$)    &  88 &      105 &      157 &       55 &       41 &       47 
\enddata
\tablenotetext{a}{Noise is measured per 1.8$\arcmin$ pixel.}
\label{tbl1}
\end{deluxetable}

\begin{deluxetable}{lcccccccccc}
\rotate
\scriptsize
\tablecaption{Multi-frequency properties of regions with high reliability excess at 143 GHz.  Note that the first two regions are effectively the same.\tablenotemark{a}}
\tablehead{
\colhead{$(l,b)$} & \colhead{$\theta$} & \colhead{Technique} & \colhead{I$_{100}$} & \colhead{SNR$_{143}$} & \colhead{I$_{143}$} & \colhead{I$_{217}$} & \colhead{SNR$_{353}$} & 
\colhead{I$_{353}$} & \colhead{I$_{545}$} & \colhead{I$_{857}$}
}
\startdata
    $(86.8, -69.4)$  & 1.8  &   SMICA &           1.5 $\pm$ 0.5  &  7.3 &  2.5 & 1.3 $\pm$ 0.7 &  -0.6 & -0.9 & -5.9 $\pm$ 2.9 & 1.5\\
    $(83.6, -69.4)$  & 3.7  &  SMICA  &           0.9  $\pm$ 0.2 &  9.2 &  1.3 & 0.3 $\pm$  0.3&  -1.5 & -1.0 & -2.7 $\pm$ 1.2 &    1.4\\\\
 
    $(93.2,  69.4)$  & 1.8   &  SMICA   &         2.2 $\pm$ 0.5  &  7.8 &   3.2 & 0.9 $\pm$  0.7 & -0.8 & -1.5 &-12.9 $\pm$ 3.6 &    1.1\\
 
   $(151.9,  31.4)$  & 1.8   & Commander &  0.8 $\pm$ 0.3 &  5.7 &   1.3 &-0.1 $\pm$  0.5 & -1.3 & -1.8 & -1.3 $\pm$ 3.1 &    2.4\\
   $(341.7, -22.0)$ &   1.8  & Commander  & 1.6 $\pm$ 0.4 &  9.0 &   2.2 & 0.5 $\pm$  0.6 & -1.4 & -1.8 & -3.6 $\pm$ 2.2  &   3.2\\
\enddata
\tablenotetext{a}{Galactic coordinates $(l,b)$ are given in degrees, while $\theta$ is the angular resolution of the rebinned maps in degrees. Intensities are given in kJy\,sr$^{-1}$ units except for I$_{857}$
which is in MJy\,sr$^{-1}$.}
\label{tbl2}
\end{deluxetable}

\begin{deluxetable}{ccccc}
\tablecaption{Significance of excess from a Monte-Carlo analysis of foregrounds}
\tablehead{
\colhead{$(l,b)$} & \colhead{P$_{5\sigma}$\tablenotemark{a}} & \colhead{P$_{3\sigma}$\tablenotemark{b}} & \colhead{$\alpha$\tablenotemark{c}} & \colhead{SNR$_{excess,143}$\tablenotemark{d}}
}
\startdata
$(83.6, -69.4)$ &  44 & 76 & -1.17 & 5.4 \\
$(86.8, -69.4)$ &  12 & 55 & -0.14 & 3.2 \\
$(93.2,  69.4)$ &  27 & 70 & -1.02 & 4.3 \\
$(151.9,  31.4)$ &  6 & 36 & -0.49 & 3.2 \\
$(341.7, -22.0)$ & 35 & 67 & -1.10 & 5.2 \\
\enddata
\tablenotetext{a}{Percent probability of the excess being significant at the $>5\sigma$ level.}
\tablenotetext{b}{Percent probability of the excess being significant at the $>3\sigma$ level.}
\tablenotetext{c}{Spectral index of the foreground emission fit to the 100 and 217 GHz residuals.}
\tablenotetext{d}{SNR$_{excess,143}$ is the signal to noise of the residual excess at 143 GHz after subtracting a power-law fit to the 100 and 217 GHz residuals from Table \ref{tbl2}.}
\label{tbl3}
\end{deluxetable}

\appendix
\counterwithin{figure}{section}

\section{Foreground Ratio Maps}

\begin{figure}
\plotone{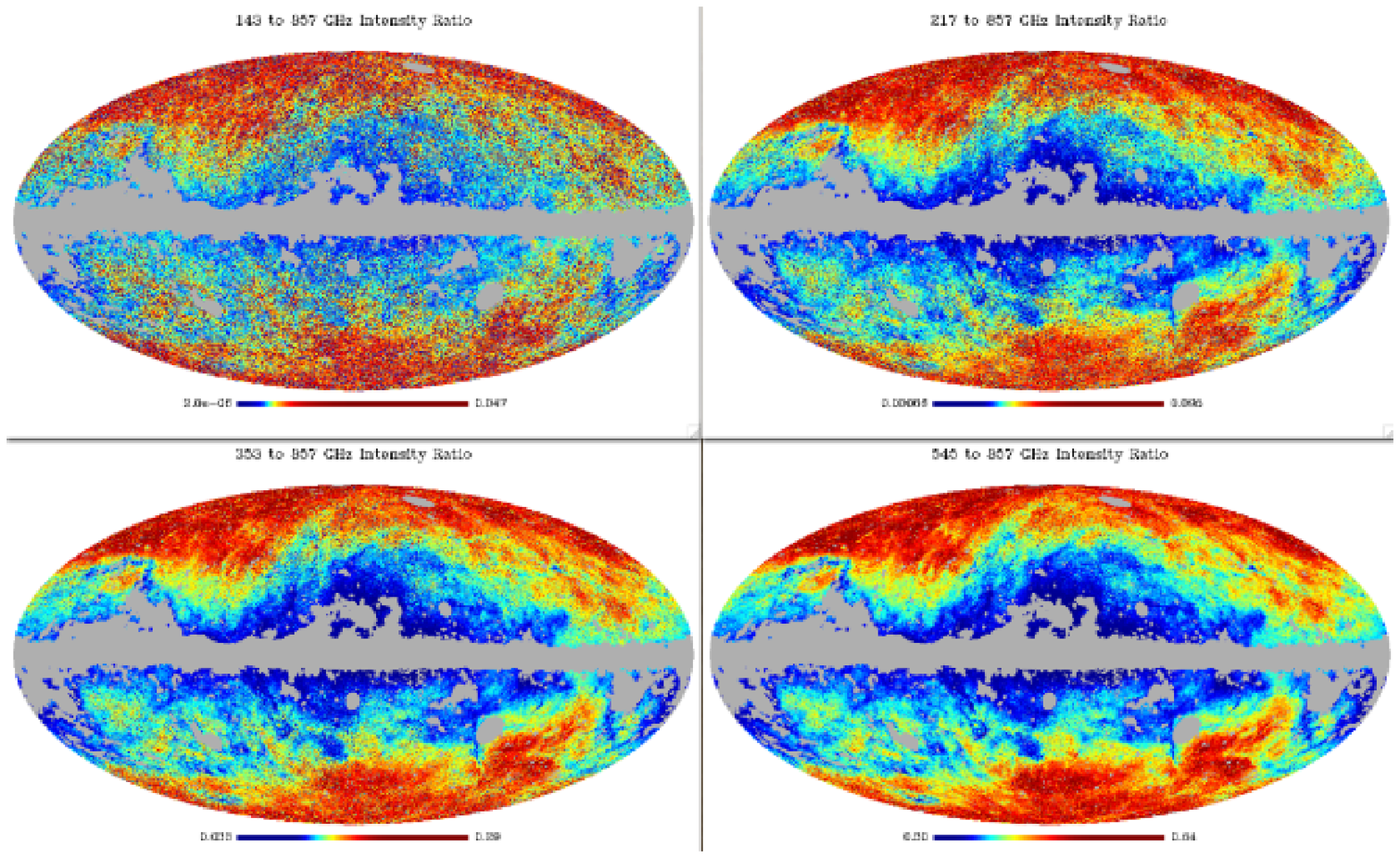}
\caption{
All-sky maps in Galactic Mollweide projection at NSIDE=2048 (1.8$\arcmin$) resolution showing the ratio of intensity at a particular frequency to the 857 GHz intensity. As can be seen,
higher Galactic latitudes tend to have cooler ISM emission and thereby higher ratios, than regions close to the Galactic Plane which has been masked as described in the text.
}
\label{fig:ratiomaps}
\end{figure}

In this Appendix, we show the intensity ratio maps of foreground emission that are generated from the CMB-free all-sky maps (Figure \ref{fig:ratiomaps}). We note that the ratios are measured
at the location of the CMB hot spots identified in the component-separated CMB map and then applied to the CMB cold spots as described in the text.

\section{Properties of Residuals at Each Frequency}

\begin{figure}
\plotone{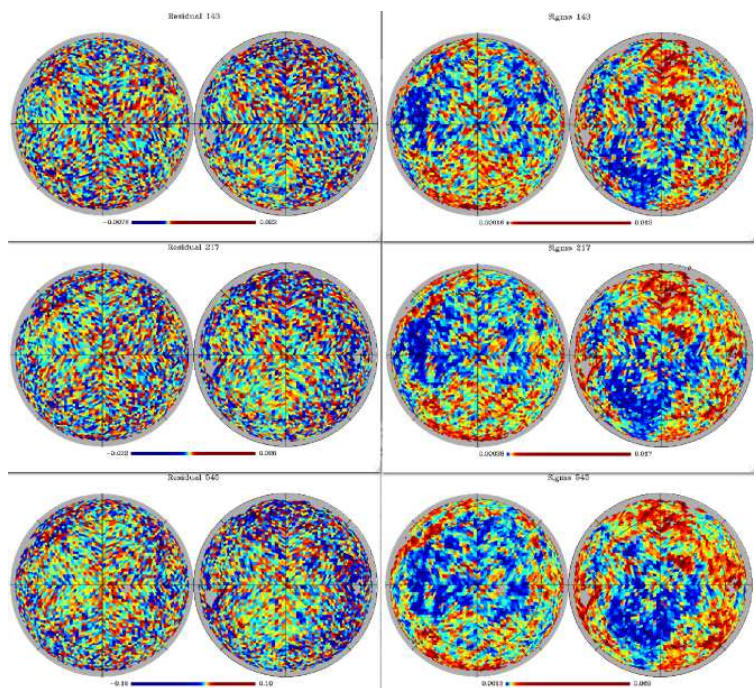}
\caption{
The left two columns shows the residual maps at 143, 217 and 545 GHz on a 1.8$\arcdeg$ pixel scale.  The maps are in orthographic projection with the
Northern hemisphere on the left and the Southern hemisphere on the right. The Galactic equator is the circumference of each circle. The Galactic center is at the bottom of the circle while the
North Galactic Pole is the center of the left circle while the South Galactic Pole is the center of the right circle.
The right two columns shows the noise maps at the same frequencies with the same convention. Units are MJy\,sr$^{-1}$. Due to the range of noise values resulting from the scan strategy as well as
the ISM, it is incorrect to use a single noise value per frequency for the analysis. Dividing the residual map with the noise map reveals regions with highly significant excess residual emission.}
\label{fig:snrmaps}
\end{figure}

\begin{figure}
\plotone{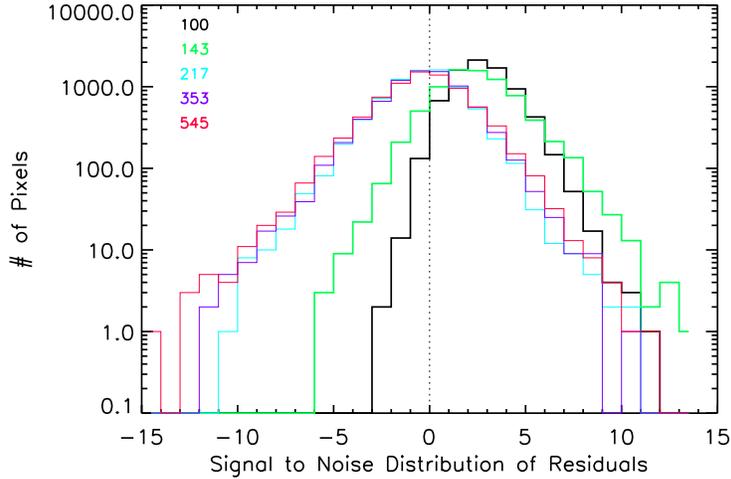}
\caption{
Histograms of the residual emission in the cleaned 1.8$\arcdeg$ resolution maps at the location of CMB cold spots at each of the \Planck\ frequencies. Only the 100 and 143 GHz bands show evidence for excess emission, with the median excess consistent with Galactic synchrotron radiation with a I$_{\nu}\propto\nu^{-0.69}$ spectrum. 
}
\label{fig:residualhist}
\end{figure}

\begin{figure}
\plotone{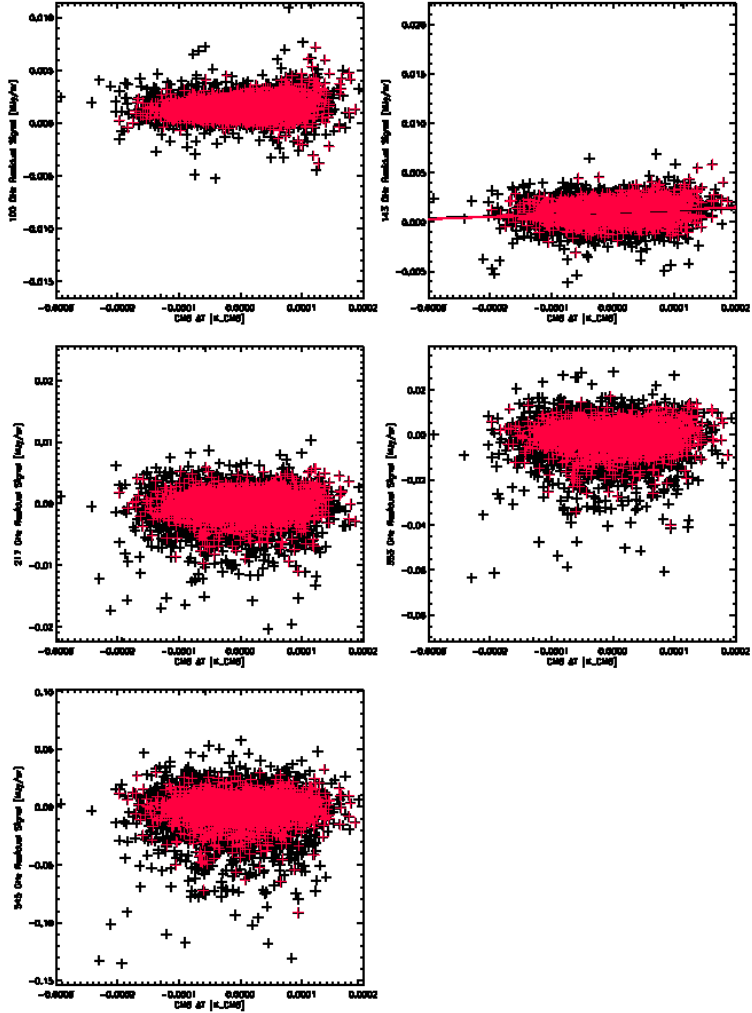}
\caption{
Residual emission at each of the frequencies shown against the CMB temperature decrement. The black points are each pixel in the 1.8$\arcdeg$ resolution maps with  ISM$<8$\,MJy/sr.
The red points are only the ones with  ISM$<4$\,MJy/sr and at Galactic latitudes greater than 45$\arcdeg$. Except for 100 and 143 GHz, the median of the residuals at each frequency is consistent with zero.
}
\label{fig:cmbcorr}
\end{figure}

\begin{figure}
\plotone{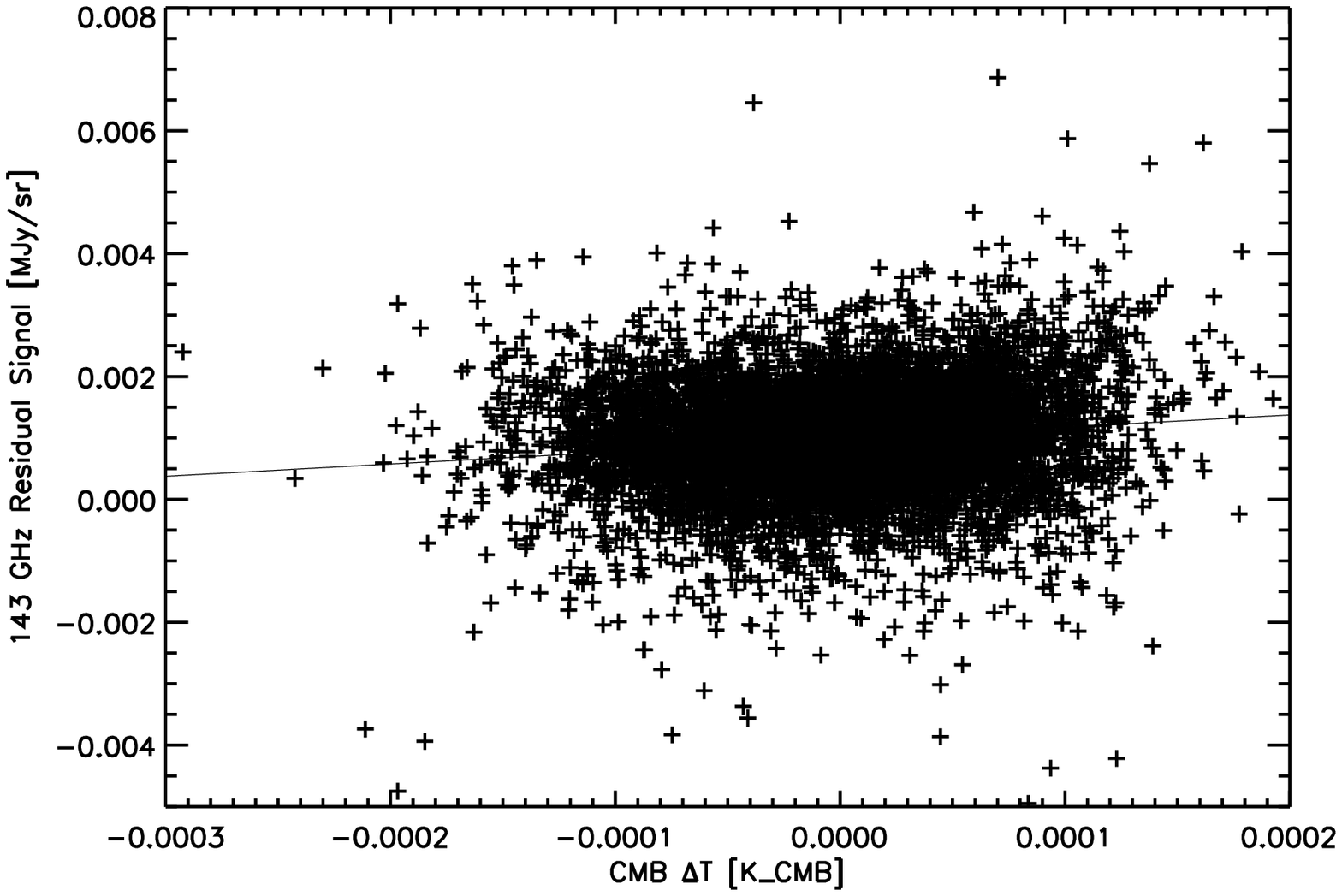}
\plotone{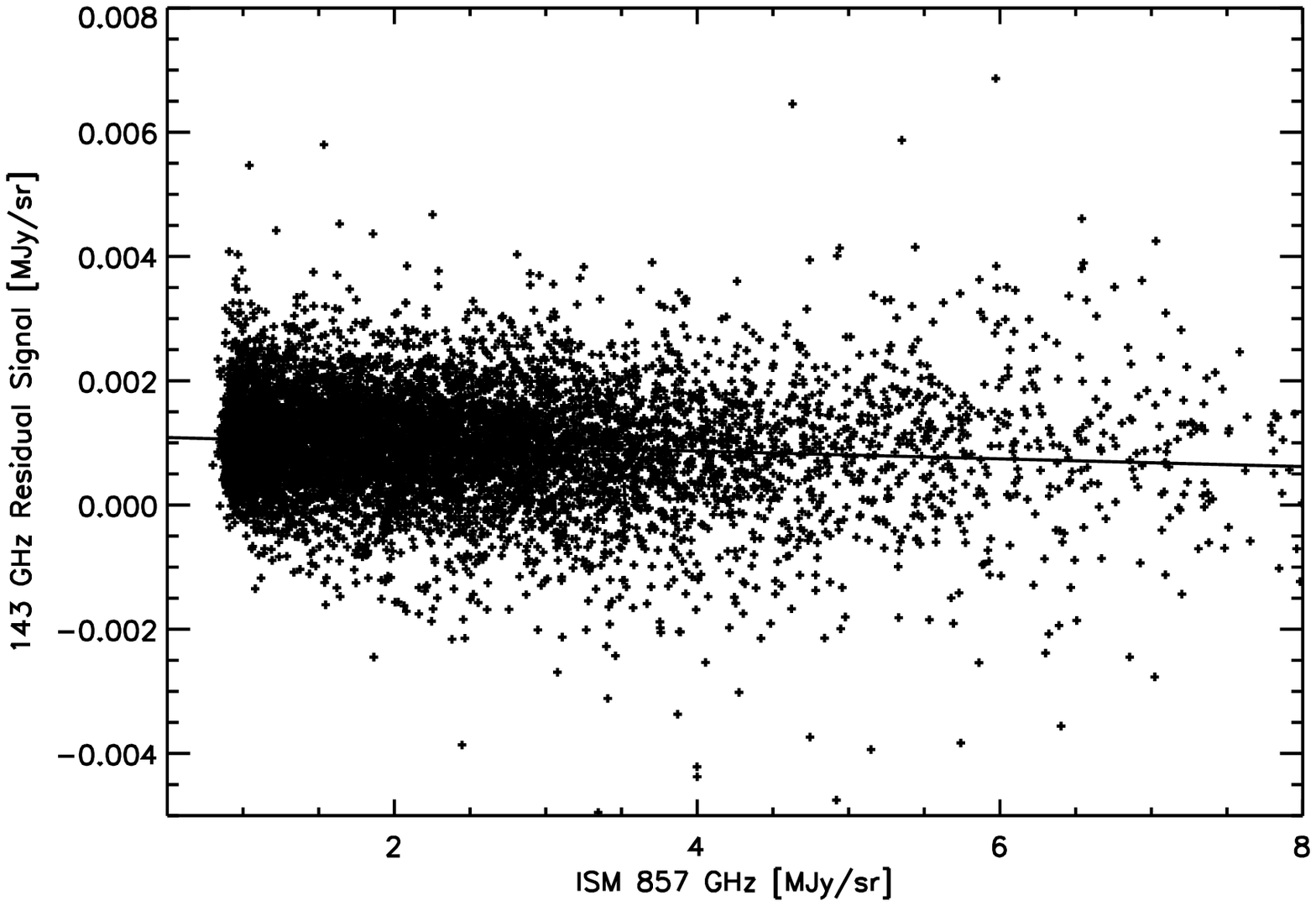}
\caption{
(Top) The strength of the 143 GHz residual signal in the 1.8$\arcdeg$ resolution maps plotted against the temperature in the correspondingly degraded
CMB map with the solid line showing the best fit.
(Bottom) The strength of the 143 GHz residual signal from CMB cold spots in the 1.8$\arcdeg$ resolution maps correlated against the 857 GHz intensity for all pixels with ISM$<8$\,MJy/sr. There is no evidence for a correlation here, indicating that the ISM has been well removed. However, as described in the text, the residual emission is dominated by synchrotron emission.
}
\label{fig:ismcorr}
\end{figure}

After cleaning the individual frequency maps of CMB and foreground emission, the maps have a residual intensity with some scatter at the location of the CMB cold spots . At the location of the hot spots, 
the residual intensity is zero by design. These high resolution maps are then re-binned to spatial resolutions of 1.8$\arcdeg$ and 3.7$\arcdeg$ which averages over the residual emission in a
 large number of CMB cold spots and reduces
the noise in the measurement (Figure \ref{fig:snrmaps}). Of all the frequencies, only 100 and 143 GHz show a clear positive bias in the residual signal with the median of the distribution located at 2$\sigma$.
The other frequencies have a median of zero (Figure \ref{fig:residualhist}). The residual intensities are uncorrelated with either the ISM emission or the CMB (Figures \ref{fig:cmbcorr} and
\ref{fig:ismcorr}). 
However, the spectrum of the residual is consistent with that of Galactic synchrotron emission.

\begin{figure}
\plotone{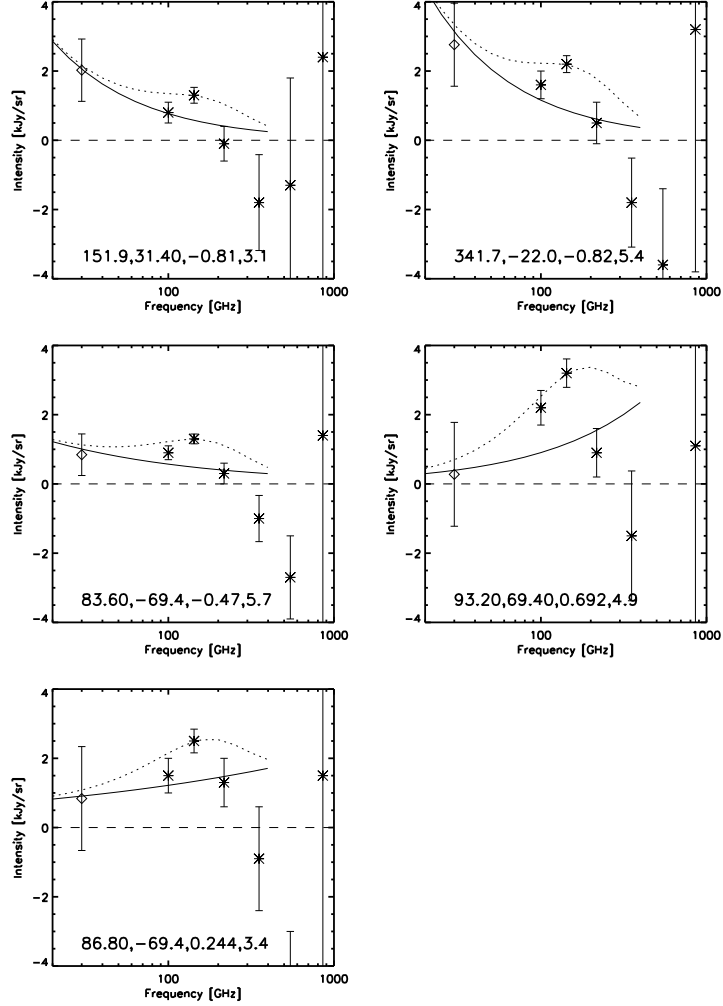}
\caption{
The five panels show as stars the spectrum of emission in the five regions in Table \ref{tbl2}. Also shown (diamond) is the residual 30 GHz intensity from the \Planck\ LFI maps of the same resolution. 
The solid line is the best fit to the 30, 100 and 217 GHz points with the four numbers at the bottom left being $(l,b,\alpha,SNR_{excess,143})$ where $\alpha$ is the spectral index of the fit and 
 SNR$_{excess,143}$ is as defined in Table \ref{tbl3}.  While the 3$\sigma$ points have a random probability of occurring a handful of times in a map with 12000 pixels, the $\gtrsim$5$\sigma$
 points should occur only once in 70 maps and the 5.7$\sigma$ point once in 70000 maps. The dotted line shows the result of adding a continuum black body component of 2.7K, which may be due
 to residual CMB or ISM; the measurements at 100 and 217 GHz are clearly violated, arguing that the excess is more likely due to line emission.
}
\label{fig:ismresid}
\end{figure}

After fitting for Galactic synchrotron emission, including 30 GHz \Planck\ data which has been similarly cleaned of CMB fluctuations and sources, we find the residuals at 143 GHz
continue to exceed the foregrounds by $3.1-5.7\sigma$ (Figure \ref{fig:ismresid}). Although the $\sim$3$\sigma$ fluctuations could occur by chance a handful of times in a map with $\sim$12000 pixels, the
$\gtrsim$5$\sigma$ excess points in 3 of the regions have a negligible chance of occurring by random. It must therefore be due to an additional component of emission as discussed in the text.

\begin{thebibliography}{11}
\expandafter\ifx\csname natexlab\endcsname\relax\def\natexlab#1{#1}\fi

\bibitem[Aguirre 
\& Tegmark(2005)]{Aguirre2005} Aguirre, A., \& Tegmark, M.\ 2005, \jcap, 1, 003 

\bibitem[Aguirre 
\& Johnson(2011)]{Aguirre2011} Aguirre, A., \& Johnson, M.~C.\ 2011, Reports on Progress in Physics, 74, 074901 

\bibitem[Ali-Ha{\"i}moud \& Hirata(2011)]{Haimoud2011} Ali-Ha{\"i}moud, Y., \& Hirata, C.~M.\ 2011, \prd, 83, 043513

\bibitem[Ali-Ha{\"i}moud(2013)]{Haimoud2013} Ali-Ha{\"i}moud, Y.\ 
2013, \prd, 87, 023526 

\bibitem[Chang et al.(2009)]{Kleban} Chang, S., Kleban, M., 
\& Levi, T.~S.\ 2009, \jcap, 4, 025 

\bibitem[Chluba \& Sunyaev(2006)]{Chluba2006} Chluba, J., \& Sunyaev, R.~A.\ 2006, \aap, 458, L29

\bibitem[Chluba et al.(2007)]{Chluba2007} Chluba, J., 
Rubi{\~n}o-Mart{\'{\i}}n, J.~A., \& Sunyaev, R.~A.\ 2007, \mnras, 374, 1310 

\bibitem[Chluba \& Sunyaev(2009)]{Chluba2009} Chluba, J., \& Sunyaev, R.~A.\ 2009, \aap, 501, 29 

\bibitem[Cormier et al.(2015)]{Cormier} Cormier, D., Madden, S.~C., Lebouteiller, V., et al.\ 2015, \aap, 578, A53 

\bibitem[Dai et al.(2013)]{Dai2013} Dai, L., Jeong, D., 
Kamionkowski, M., \& Chluba, J.\ 2013, \prd, 87, 123005 

\bibitem[Desjacques et al.(2015)]{Desjacques2015} Desjacques, V., Chluba, J., Silk, J., de Bernardis, F., \& Dor{\'e}, O.\ 2015, arXiv:1503.05589

\bibitem[Feeney et al.(2011)]{Feeney2011} Feeney, S.~M., Johnson, 
M.~C., Mortlock, D.~J., \& Peiris, H.~V.\ 2011, \prd, 84, 043507 

\bibitem[G{\'o}rski et al.(2005)]{Healpix} G{\'o}rski, K.~M.,  Hivon, E., Banday, A.~J., et al.\ 2005, \apj, 622, 759 

\bibitem[Hansen et al.(2009)]{Hansen2009} Hansen, F.~K., Banday, 
A.~J., G{\'o}rski, K.~M., Eriksen, H.~K., 
\& Lilje, P.~B.\ 2009, \apj, 704, 1448 

\bibitem[Kogut et al.(2014)]{Kogut} Kogut, A., Chuss, D.~T., 
Dotson, J., et al.\ 2014, \procspie, 9143, 91431E 

\bibitem[Lamarre et al.(2010)]{Lamarre2010} Lamarre, J.-M., Puget, J.-L., Ade, P.~A.~R., et al.\ 2010, \aap, 520, A9

\bibitem[Mather et al.(1994)]{Mather1994} Mather, J.~C., Cheng, E.~S., Cottingham, D.~A., et al.\ 1994, \apj, 420, 439 

\bibitem[{\"O}berg et al.(2010)]{Oberg2010} {\"O}berg, K.~I., 
Bottinelli, S., J{\o}rgensen, J.~K.,  \& van Dishoeck, E.~F.\ 2010, \apj, 716, 825 

\bibitem[Oesch et al.(2015)]{Oesch2015} Oesch, P.~A., Bouwens, 
R.~J., Illingworth, G.~D., et al.\ 2015, \apj, 808, 104 

\bibitem[Planck Collaboration et al.(2014a)]{Overview2014} Planck Collaboration, et al.\ 2014, \aap, 571, A1

\bibitem[Planck Collaboration et al.(2014b)]{HFIBandpasses} Planck Collaboration, et al.\ 2014, \aap, 571, A9      

\bibitem[Planck Collaboration et al.(2014c)]{HFICO} Planck Collaboration, et al.\ 2014, \aap, 571, A13              

\bibitem[Planck Collaboration et al.(2015)]{Isotropy2015} Planck 
Collaboration, et al.\ 2015, arXiv:1506.07135

\bibitem[Planck Collaboration et al.(2015)]{Calib2015} Planck 
Collaboration, et al.\ 2015, \aap, submitted, arXiv:1502.01587 

\bibitem[Planck Collaboration et al.(2015)]{CompSep2015} Planck 
Collaboration, et al.\ 2015, \aap, submitted, arXiv:1502.01588 

\bibitem[Planck Collaboration et al.(2015)]{Parameters2015} Planck 
Collaboration, et al.\ 2015, arXiv:1502.01589 

\bibitem[Planck Collaboration et al.(2015)]{CMBmaps2015} Planck 
Collaboration, et al.\ 2015, \aap, submitted, arXiv:1502.05956 

\bibitem[Planck Collaboration et al.(2015)]{PCCS2} Planck 
Collaboration, et al.\ 2015, A\&A, submitted

\bibitem[Planck Collaboration et 
al.(2014)]{PIP15} Planck Collaboration, Ade, P.~A.~R., Aghanim, N., et al.\ 2014, \aap, 565, A103 

\bibitem[Rubi{\~n}o-Mart{\'{\i}}n et al.(2008)]{JARM2008} Rubi{\~n}o-Mart{\'{\i}}n, J.~A., Chluba, J., \& Sunyaev, R.~A.\ 2008, \aap, 485, 377

\bibitem[Sachs  \& Wolfe(1967)]{SachsWolfe67} Sachs, R.~K., \& Wolfe, A.~M.\ 1967, \apj, 147, 73 


\end{thebibliography}
\end{document}